\documentstyle[eqsecnum,prb,aps,epsf]{revtex}

\begin{document}

\draft
\preprint{}
\title{Instability of the marginal commutative model of tunneling
centers interacting with metallic environment: Role of the electron-hole
symmetry breaking}
\author{A. Zawadowski and G. Zar\'and }
\address{Institute of Physics, Technical University of Budapest,\\
H 1521 Budafoki \'ut 8., Budapest, Hungary.}
\author{P. Nozi\`eres}
\address{Institute Laue-Langevin, 38042 Grenoble Cedex, France}
\author{K.\ Vlad\'ar}
\address{Research Institute for Solid State Physics,
Budapest, P.O.\ Box 49, H-1525, Hungary}
\author{G. T. Zim\'anyi}
\address{Physics Department, University of California, Davis,
California 95616}
\maketitle

\begin{abstract}
The role of the electron-hole symmetry breaking is
investigated for a symmetrical commutative two-level system in a
metal using the multiplicative renormalization group in a
straightforward way. The role of the symmetries of the model 
and the path integral technique are also discussed in detail.
It is shown that the electron-hole symmetry breaking may 
make the model non-commutative and
generate the assisted tunneling process which is, however, too small itself
to drive the system into the vicinity of the two-channel Kondo fixed
point. While these results are in qualitative agreement with
those of Moustakas and Fisher (Phys. Rev. B {\bf 51}, 6908 (1995),
{\it ibid} {\bf 53}, 4300 (1996)) the scaling equations turn out to
be essentially different. We show that the main reason for this
difference is that the procedure for the elimination of the high
energy degrees of freedom used by Moustakas and Fisher leaves only the
free energy invariant, however, the couplings generated are not
connected to the dynamical properties in a straightforward way
and should be interpreted with care. These latter results might have
important consequences in other cases where the path integral
technique is used to produce the scaling equations and calculate
physical quantities.
\end{abstract}
\pacs{PACS numbers: 72.10.Fk, 72.15.Cz, 71.55.-i}

\eject
\setlength{\textheight}{19cm}

\section{Introduction}
\label{sec:introduction}
\setlength{\textheight}{19cm}

The two-channel Kondo model originally introduced for magnetic
ions with crystalline splitting \cite{Noz_Bland,Cox} and tunneling
two-level systems (TLS) in metals \cite{Zawa} has recently
attracted broad interest.\cite{Ralph,Review} In some of those problems the orbital variables
play the role of the spins of the impurity and the conduction electron
while the real spins of the conduction electrons are not
dynamical variables but instead represent an extra twofold
degeneracy for the conduction electrons interacting with the
impurity. A TLS can be considered as a good realization of the two-channel
orbital Kondo problem.\cite{Zawa,Ralph,Review,Vlad_Zaw,Zar_Vlad,JvD}
Besides the fast TLS's and the impurity ions with orbital
dynamical variables mentioned before the two-channel Kondo
problem can also be related to the two-impurity Kondo
problem.\cite{Jones_Varma,Affl_Jones} Due
to the double degeneracy of the real conduction electron spins
the fixed point of these models is in the intermediate strong coupling
region.\cite{Noz_Bland}


It became obvious in recent years that the breaking of
electron-hole (e-h) symmetry may result in the instability of the
fixed points found in the e-h symmetric case,\cite{Silva,Kunose}
and new fixed points  might become relevant as in the two-impurity Kondo
problem.\cite{Jones_Varma,Affl_Jones}

The present paper is devoted to a special class of the TLS
problem, called the {\it dissipative} or {\it commutative} TLS
model,\cite{BVZ,Leggett,KaganProkofev,Costi} where there is only
one dynamical coupling to describe the electron-TLS interaction in
contrary to the orbital Kondo problem ({\it non-commutative} TLS
problem),  where there must be at least two non-commuting interaction terms. 
Usually, when dealing with such impurity models an e-h
symmetrical conduction electron band is assumed. Then the
dissipative two-state model provides an example of marginal models in the
renormalization group sense, i.e., its dynamical couplings
remain unrenormalized under scaling. The main goal of the
present paper is to show that 
the commutative model becomes unstable due to  the
presence of e-h symmetry breaking of a specific form,
and serves as a toy model to examine the role of e-h symmetry breaking
occurring in a realistic model of a TLS.
As we shall see, in the presence of e-h symmetry the model
has an additional symmetry which ensures the stability of the
line of marginal fixed points. This problem has its own interest even if
it does not play a major physical role in most of the cases of realistic
systems. By developing numerical estimates, we will argue that 
for realistic parameters of a TLS, this effect is extremely
small for metals, and can be safely neglected for practical
purposes.  Therefore, in case the screening interaction is not
anomalously strong, the dominant mechanism making the commutative TLS
model unstable will be the {\it intrinsic assisted tunneling}
considered in Refs. \onlinecite{Zawa} and \onlinecite{Vlad_Zaw}.

The essential effect of the  e-h symmetry breaking can be
demonstrated by considering the two time-ordered diagrams in
Fig.~\ref{fig:second}. If the conduction
electron is scattered by a simple potential scatterer $V$,
both diagrams provide a diverging term $\sim V^2 \ln{D\over
\omega}$, ($D$ and $\omega$ being the conduction electron
bandwidth and some small-energy scale, respectively), but these
contributions cancel each other due to e-h symmetry.\cite{BVZ} That is
one of the most important ingredients of the X-ray absorbtion
problem.\cite{Roulet} In case of e-h symmetry breaking the
cancellation is not exact, and a non-diverging contribution
remains. Such contribution may change the mathematical structure
and the universality class of  the theory.

 To be specific, we imagine a TLS, where a heavy particle (HP) can
jump between two positions in a double potential well.
In general the HP can dynamically interact with the conduction
electrons in two different ways (see Fig.~\ref{fig:potential}):

($i$) The electrons try to form a screening cloud around the
heavy particle.  
If only this interaction is taken into account then, 
using the logarithmic approximation, the usual e-h symmetrical
model\cite{Vlad_Zaw} 
(called the dissipative or commutative TLS model mentioned above) 
results in a marginal theory, 
where the screening interaction is unrenormalized but the
hopping rate of the HP is essentially reduced because the
overlap matrix element of the electronic screening clouds 
at  neighboring positions of the HP vanishes.\cite{orthogonality}

($ii$) The conduction electrons also assist the hopping
of the HP (assisted hopping) both in the case of tunneling and virtual
transitions through the excited states.\cite{ZarZaw} If the
second process ($ii$) is also taken into account then the
couplings are not commuting in the momentum space, and the
non-commutative model belonging to the class of two-channel
Kondo models is recovered. 

In the present paper we concentrate on case ($i$), where the
assisted tunneling ($ii$) is absent in the bare hamiltonian.
As mentioned above, in the original treatment of the problem
one usually assumes that the local density of states is the same
for each orbital channel of the conduction electrons and that it
is e-h symmetrical. In a realistic case, however, these
assumptions are not a priori justified. 


Recently, using a path integral approach Moustakas and Fisher
\cite{MF1,MF2} argued that in the commutative problem ($i$)
the above mentioned potential scattering at the HP is relevant
and changes the scaling behavior of the TLS.  The possibility
has been pointed out that such a potential may generate an
electron assisted hopping. Such a potential is usually neglected
which is certainly not justified for the case when the heavy
particle is different from the atoms of the host material. (The
problem of this static potential scattering has been
investigated earlier by Kagan and Prokof'ev by applying the
adiabatic renormalization technique.\cite{KaganProkofev})

As we shall see both scattering on a local static potential and
a realistic dispersion relation for the conduction electrons
validate the previous assumptions for the local density of
states, and generate process ($ii$).  Therefore, in the present
paper we shall drop them and treat a general local density of
states incorporatring both mechanisms.  

To investigate the relevance of the static potential and the
generation of assisted tunneling two different methods
 will be developed. First, applying a straightforward multiplicative
renormalization group method in the leading and next to leading
logarithmic approximation, 
appropriate in the small coupling limit,
we show that while the energy
dependence of the local electronic spectral functions really
generates the assisted tunneling term, it can  be
neglected in most of the cases and the commutative behavior is
recovered. We also 
investigate the problem with a path integral technique. In the
concluding section of this paper and Appendix B we discuss the
ambiguity in the construction of the scaling equations derived
by Moustakas and Fisher.\cite{MF1,MF2} We point out that in the
path integral approach the elimination of the events with small
time difference is not uniquely defined, and two procedures are
possible: ($i$) Any events being closer to each other than a
certain time difference must be replaced by a generated
interaction \cite{MF1}, ($ii$) Only those  processes must be
eliminated, where electrons are involved. While both procedures
leave the free energy of the HP unchanged, the couplings
generated and the electronic sector behave essentially
differently. We will argue that the physical argument
behind the second procedure is more sound.

It is interesting to note that the first method applied by
Moustakas and Fisher generates the electron-TLS coupling even in
that trivial case where there is {\it no dynamical coupling}
between the TLS and the conduction electrons.\cite{MF1,MF2} 
(This importaint point has also been mentioned in footnote 20 of
Ref.~\onlinecite{MF2}). 
We show that the meaning and the application of these couplings to 
calculate real physical quantities is not straightforward, as
the scaled couplings are usually plugged into formulas obtained
by conventional diagram technique not directly related to the
path integral method.  In this way different sets of the scaled
couplings lead to different measurable quantities. Thus in the
path integral formalism no more than one if any scaling
procedure can be appropriate to provide the correct scaled
couplings to calculate physical quantities. 
Since the path integral technique should be also correct in the
weak coupling region where the multiplicative renormalization
group technique is adequate, we can decide which procedure is
correct by comparing the scaling equations in this limit.
We show that the
second scaling procedure is the one which reproduces the
dynamical behavior obtained in the next to leading logarithmic
approximation. With the elimination procedure $(ii)$ the
assisted hopping is {\it not generated} by the combination of
spontaneous tunneling and simple potential scattering if all the
dynamical couplings are switched off.

We also show using the multiplicative renormalization group method 
that in the presence of e-h symmetry and without
assisted tunneling the TLS-conduction electron dynamical couplings remain
unrenormalized  up to the next to leading logarithmic order.
This also contradicts to the elinimation scheme used in
Refs.~\onlinecite{MF1,MF2}. 

The relevance of different terms in the Hamiltonian has been
discussed by Moustakas and Fisher in terms of symmetry
breaking.\cite{MF1,MF2}  In order to clarify
the problem special emphasis will be put on these symmetries.  We show
that the conventional commutative TLS model 
without potential scattering and with an e-h symmetrical band
has an additional symmetry reflecting the role of e-h symmetry. 
This symmetry is a combination of  reflection to the middle
of the TLS for the conduction electrons
(but not for the tunneling HP), and the e-h symmetry, and it can be broken
in different ways, e.g. by potential scattering, an asymmetrical
conduction electron band, or an energy dependent coupling
between the HP and the conduction electrons.  
(This latter, analized in Appendix~B of Ref.~\onlinecite{MF1} is
closely related to the energy dependence of the local density of
states, studied in the present paper.)
in the Hamiltonian which do not conserve the number of right
(left) electrons in the sense introduced by Moustakas and
Fisher.\cite{MF1} The breaking of e-h symmetry can also be the
consequence of an energy dependent  coupling in
agreement with the suggestion of Moustakas and
Fisher.\cite{MF1,MF2} The new symmetry found guarantees that the
low energy fixed point is marginal (the Hamiltonian is invariant
under the scaling procedure) and no Kondo effect occurs.  On the
other hand if it is violated then in general the assisted
tunneling is generated and the system starts to flow 
in the direction of 
the two-channel Kondo fixed point until the energy scale becomes
smaller than the renormalized splitting between the two lowest
states of the TLS, and the TLS dynamics is frozen out.

However, we shall see that the assisted tunneling generated in this way is
very weak  and usually unobservable. In fact, the numerical
renormalization group study of the same problem by Libero and
Oliveira in the absence of e-h symmetry
does not indicate the scaling toward the intermediate coupling
fixed point.\cite{LibOliv} 
This is due to the fact that since the generated assisted tunneling is
always very weak the splitting of the TLS stops the
scaling before the system could approach the neighborhood of the
two-channel Kondo fixed point. It will be speculated, that it can result
in an observable effect only in the case of very asymmetrical  band
(degenerate semiconductors).\cite{Fukuyama}

In general the question can be raised whether a spontaneous
hopping process renormalized by the screening interaction
between the TLS and the electrons can induce an effective
assisted hopping or not.
A straightforward calculation of the diagrams in
Fig.~\ref{fig:maindiag} shows that electron assisted hopping is
generated but only if the local electronic propagator breaks e-h
symmetry. The simplest realization of such symmetry breaking is
an asymmetric electron band or a static potential scattering at
the tunneling centers.  The diagram in Fig.~\ref{fig:maindiag} is
logarithmic, but it contains an extra small factor
$\sim\Delta_0/\epsilon_F \sim 10^{-4}$, $\Delta_0$ being the
amplitude of spontaneous tunneling of the TLS and $\epsilon_F$
the Fermi energy. Such processes are usually dropped in the next
to leading logarithmic calculations because of the appearance of
the small term $\Delta_0/D$ in these equations. The appearance
of this instability can be traced back in the path integral
technique and supports the legitimity of the the second
elimination scheme discussed above.  

The paper is organized as follows. In Sec. II the model is
introduced and the time-ordered perturbation theory is applied to
demonstrate the importance of the e-h symmetry breaking and how
the assisted tunneling is generated. In Sec. III the additional
symmetry is introduced in the presence of e-h symmetry and it is
shown how that insures the stability of the marginal fixed
point. In Sec. IV the strength of the assisted tunneling
generated by the breaking of the e-h and the additional symmetry
is estimated. In Sec. V the results and the ambiguity in the
path integral method are discussed. In Appendices A and B
we calculate  the amplitude of the different e-h symmetry breaking
parameters and {\it e} show how the scaling equations obtained in the different path
integral techniques may be related by a formal transformation.
Finally, in Appendix C and D we investigate the mixing of the high-
and low-energy conduction electron degrees of freedom due to a
simple potential scatterer and the e-h symmetry in the case of a 
cubic tight binding model.

\section{The model and application of the time ordered diagram
technique}
\label{sec:timeord}

First we introduce the model and the notations used closely
following Ref.~\onlinecite{MF2}. 
Since the results of this Section are independent of the real
spin indices of the electrons, for the sake of simplicity we
drop them throughout  this Section.  
The creation operators of the HP at
site 1 and 2 are denoted by $d^+_1$ and $d^+_2$.
The conduction electron annihilation operators at positions $\pm
{\bf R}/2$, $c_{\pm}$,
can be expressed in terms of the annihilation operators for
spherical waves (s-waves) with momentum $k$ at the positions
${\bf r}=\pm{\bf  R}/2$  denoted by $c_{+,k}$ and $c_{-,k}$:
\begin{equation}
c_{\pm}= \int {dk\over 2\pi} c_{\pm
k}=\int{dk\over2\pi}\int{k^2d\Omega_{\bf k}\over(2\pi)^2} 
\exp(\pm i{\bf k}{\bf R}/2) c_{\bf k} \;,
\label{eq:c_pm}
\end{equation}
where the $c_{\bf k}$'s denote plane wave annihilation
operators. 
Furthermore for convenience the following even and odd operators
 are defined:
\begin{eqnarray}
c_{ek}^+ &=&{1\over \sqrt{N_e(k)}}(c_{+,k}^+ + c_{-,k}^+)\;,
\nonumber \\
c_{ok}^+ &=& {1\over \sqrt{N_o(k)}}(c_{+,k}^+ - c_{-,k}^+)\;,
\label{eq:c_e,o}
\end{eqnarray}
where the normalization constants are given by
\begin{equation}
N_{e,o} (k) = {2 k^2 \over \pi} \left( 1 \pm {\sin kR \over
kR}\right) \;,
\label{eq:normalization}
\end{equation}
and the operators in Eq.~(\ref{eq:c_e,o}) satisfy the anticommutation
relations: $\{c_{\alpha k},c^+_{\beta k^\prime} \} = 2\pi
\delta(k-k^\prime)\delta_{\alpha\beta}$ with $\alpha,\beta=e,o$.
The spin variables are not indicated in the above formulae. Then
the Hamiltonian can be written in a concise form in terms of the
operators
\begin{equation}
c_{e(o)}^+ = \int {dk\over 2\pi} c_{e(o),k}^+ \;.
\label{eq:c_(e,o)}
\end{equation}
Assuming a local HP-electron interaction, the Hamiltonian
considered by Moustakas and Fisher can be written as
\cite{MF1,MF2} 
\begin{equation}
H= H_0 + H_d + U \; ,
\label{eq:Hamilt}
\end{equation}
where
\begin{eqnarray}
H_0 &=& \int {dk\over 2\pi} \epsilon_k (c^+_{ek}c_{ek} +
c^+_{ok}c_{ok}) \;,\\
H_d &=& \Delta_0 (d^+_1 d_2 + d^+_2 d_1 ) \; , \\
U &=& V_1(c^+_e c_e + c^+_o c_o) + V_2 (c^+_e c_e - c^+_o c_o)
+ V_3 (d^+_1 d_1 - d^+_2 d_2 ) (c^+_e c_o + c^+_o c_e) \;.
\label{eq:int}
\end{eqnarray}
Here $\epsilon_k$ is the energy of the conduction electrons,
$\Delta_0$ is the spontaneous tunneling rate of the HP, and the
different momentum dependent couplings have been approximated
by their values at the Fermi energy.
The terms proportional to $V_1$ and $V_2$ describe
potential scatterings while $V_3$ denotes the amplitude of the
screening interaction. Note that  in the interaction part $U$ the
constraint $d^+_1 d_1 + d^+_2 d_2=1$ has already been taken into
account. In the case of an asymmetric double
potential well there is an additional asymmetry term for the HP
\begin{equation}
H_d^\prime = \Delta^z (d_1^+ d_1 - d_2^+ d_2)\;.
\end{equation}

The couplings $V_1$ and $V_2$ can be incorporated to the
one-particle properties of the conduction electrons. The
unperturbed one particle Green's functions are usually defined
as
\begin{equation}
G_{ek}^{(0)} (\omega_n) = G_{ok}^{(0)}(\omega_n) = {1\over
i\omega_n -\epsilon_k} \;,
\end{equation}
where the $\omega_n$'s denote Matsubara frequencies. Then the
electronic local Green's functions modified by the
potential scattering
\begin{equation}
\delta_{\alpha\beta}\; G_{\alpha}(\omega_n) = - \langle  T \{
c_{\alpha} c^+_{\beta}\} \rangle_{\omega_n}
\end{equation}
with $\alpha=e,o$ can be calculated easily and one obtains
\begin{equation}
G_{e,o}(\omega_n)= {G^{(0)}_{e,o}(\omega_n) \over
1 - (V_1 \pm V_2) G^{(0)}_{e,o}(\omega_n)} \;,
\end{equation}
where $G^{(0)}_{e,o}(i\omega) = \int dk\;(i\omega - \epsilon(k))^{-1}
/ 2\pi$.
The corresponding spectral functions are given by
\begin{equation}
\varrho_{e,o} ={1\over \pi}
{{\rm Im} G_{e,o}^{(0)}(\omega-i\delta) \over
\left[1-(V_1 \pm V_2) {\rm Re} G_{e,o}^{(0)}(\omega-i\delta)\right]^2
+ \left[(V_1 \pm V_2){\rm Im} G_{e,o}^{(0)}(\omega-i\delta)\right]^2}
\;.
\label{eq:spfunc}
\end{equation}
Since to the lowest order in $\omega$ the imaginary and real
parts of the Green's functions can be approximated as ${\rm Im}
G_{e,o}^{(0)}(\omega-i\delta)\sim {\rm const}$ and
${\rm Re} G_{e,o}^{(0)}(\omega-i\delta) \sim \omega$,
from Eq.~(\ref{eq:spfunc}) immediately follows that in the
presence of $V_1$ or $V_2$ the e-h 
symmetry can not hold even if the original band is symmetrical.

Now it is useful to introduce the Pauli operators
\begin{eqnarray}
d_2^+ d_2 - d^+_1 d_1 = \tau_z \;,\\
d_2^+ d_1 + d^+_1 d_2 = \tau_x
\end{eqnarray}
to describe the motion of the TLS, then  the remaining terms to be treated are
\begin{equation}
V_3 (c_e^+ c_o +c^+_o c_e) \tau_z + \Delta_0 \tau_x \;.
\end{equation}
To make the analogy with the two-channel Kondo model more
transparent 
a new rotated representation of spin operators can be
introduced, $\tau_x \rightarrow \tilde\tau_z$ and
$\tau_z \rightarrow -\tilde\tau_x$ and then the interaction
Hamiltonian becomes
\begin{equation}
- V_3 (c_e^+ c_o +c^+_o c_e) \tilde \tau_x + \Delta_0 \tilde \tau_z \;.
\end{equation}
That interaction has resemblance to the anisotropical
Kondo Hamiltonian in an external field $\Delta_0$  with a single coupling
$J_x$, but with $\varrho_e(\omega) \not = \varrho_o(\omega)$.

As we have shown above the effect of the the scattering
processes $V_1$ and $V_2$ can be incorporated into the spectral
functions $\varrho_{e,o}(\omega)$, so in the following we
formulate everything in terms of these quantities and drop the
terms $V_1$ and $V_2$. Since these are slowly
varying functions, as a first step, they can be approximated by
their values at the Fermi energy. Then treating $V_3$
perturbatively one can notice that in the different diagrams
only the combination $(\varrho_e(0)
\varrho_o(0))^{1/2}$ occurs and therefore the corresponding dimensionless
coupling is given by
\begin{equation}
g = (\varrho_e(0) \varrho_o(0))^{1/2} V_3\;.
\end{equation}
Apart from the present modification in this approximation the
calculation is identical to the one in Ref.~\onlinecite{BVZ}
Calculating the leading and the next to leading logarithmic
diagrams no corrections to the invariant couplings occur as the
second order vertex corrections cancel the self-energy correction
part. The only relevant renormalization is due to the HP wave
function renormalization which generates the renormalization of
the spontaneous splitting: $\Delta_0(\omega) = \Delta_0 (\omega
/ D)^{N_s g^2/2} $, where $D$ is a high energy cutoff of the
order of the Fermi energy, $\epsilon_F$, and
$N_s=2$ denotes the spin degeneracy of the conduction electrons.\cite{BVZ}

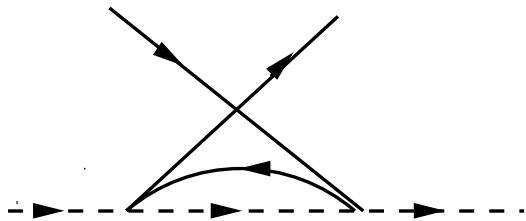
\begin{figure}
\epsfxsize=7cm
\hskip0.5cm
\vskip0.7truecm
\caption{The two first order vertex corrections for the
heavy particle (dashed line) and conduction electron (solid
line) interaction.  In diagrams (i) and (ii) the intermediate
state contains an electron and a hole, respectively. }
  \label{fig:second}
\end{figure}

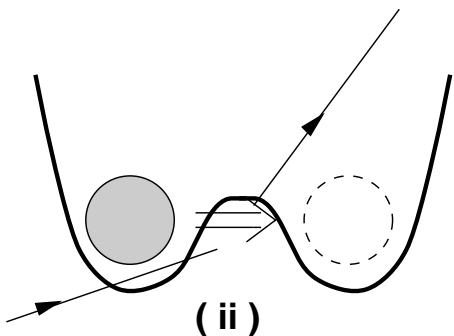
\begin{figure}
\epsfxsize=7cm
\hskip0.5cm
\vskip0.7truecm
\caption{Schematic representation of the interaction processes
between the conduction electron and the HP. In the screening
process (i) the atom sitting in one of the wells scatters the
electron while in the assisted tunneling process (ii) the
electron scattering induces the jump of the atom between the two
wells.}
\label{fig:potential}
\end{figure}

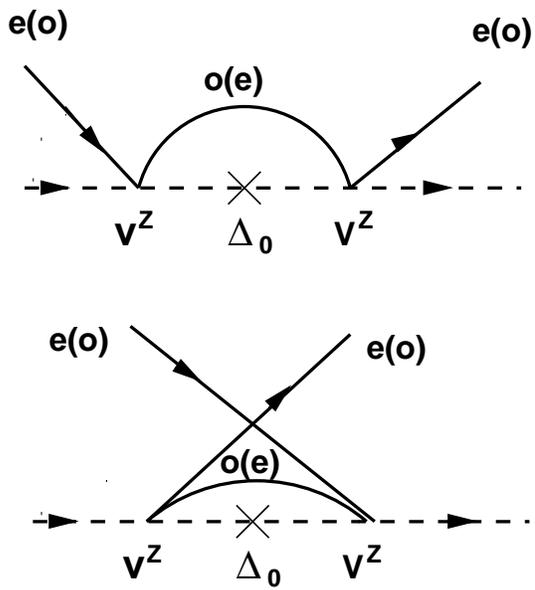
\begin{figure}
\epsfxsize=7cm
\hskip0.5cm
\vskip0.7truecm
\caption{Time ordered diagrams generating the assisted tunneling process in
the presence of appropriate e-h symmetry breaking. The
notations are identical to those of Fig.~1. The cross indicates
a spontaneous tunneling between the two positions of the HP. The
labels on the electron lines refer to the even and odd parity
electron channels with respect to reflection through the center
of the system.}
\label{fig:maindiag}
\end{figure}

The next question  is whether  the $\omega$ dependence of
$\varrho_{e,o}(\omega)$ due to the breaking of e-h
symmetry is crucial in these considerations or not. As it can be seen
from Fig.~\ref{fig:second} the diagrams for first order vertex corrections
are diagonal both in the even-odd electron channels and the TLS indices, and therefore
they only renormalize $V_{1,2}$ and $\varrho_{e(o)}(\omega)$.
Their contribution can be estimated by approximating
\begin{equation}
\varrho_{e,o}(\omega)=
\varrho_{e,o}(0)(1 + \alpha_{e,o} \omega)\;,
\label{eq:alpha}
\end{equation}
where $\alpha_{e} \not = \alpha_o \approx 0.1\;\epsilon_F^{-1} -
0.3\;\epsilon_F^{-1}$.  
Then the logarithmic part of the diagrams in
Fig.~\ref{fig:second} is exactly canceled and they give only a
constant renormalization of the potential scattering, which can
be taken into account by the redefinition of the $\alpha_{e,o}$'s.\cite{cikk_2}
The order of magnitude of this renormalization can be estimated  as
\begin{equation}
\delta (\varrho(0)V_{1,2}) \sim g^2 \alpha D \ll 1\;.
\end{equation}

It is important to note that the constants $\alpha_{e,o}$
characterize the spectrum of the local low energy excitations
coupled to the TLS, and they should remain unscaled under the
elimination of the high energy states $D \to D^\prime$. 
This subtility turns out to be non-trivial if one tries to simulate
the e-h symmetry breaking with a potential scattering term as in
Refs.~\onlinecite{MF1,MF2} since then $\alpha\sim 1/D$ (see
Appendix~C). The main difficulty  arises from the
fact that the potential scattering strongly mixes  the
original low- and high-energy excitations. One could, in
principle, proceed in two different ways: ($a$) One tries to treat
the potential scattering similarly to the dynamical interaction
terms and eliminates the high-energy degrees of the model without potential
scattering ($D\to D^\prime$). This procedure has been used in
Refs.~\onlinecite{MF1,MF2}. Since $\alpha \sim 1/D$ this
procedure obviously changes the local low-energy excitation
spectrum coupled to the TLS. ($b$) One first diagonalizes
the exactly solvable potential scattering part of the
Hamiltonian to find the real low- and high-energy excitations coupled to
the TLS and then eliminates the latter ones. This is the one
followed throughout this paper. 
Some insight to the difference between these two procedures is
given in Appendix~C.
As shown in Appendix~C, due to
the strong mixing of the original low-energy and high-energy
degrees of freedom, procedure (a) also eliminates 
a fraction of the {\it real} low energy excitations (in the
presence of potential scattering) coupled to the TLS,
and therefore the  correct procedure is the second one. 
The essential difference between the two elimination procedures
is closely related to the fact that the mass terms (i.e. the
potential scattering in our case) can never be treated at equal
footing with the interaction terms in a renormalizable field
theory.\cite{Itzykson}
We also mention that elimination scheme ($a$) results in scaling
equations which do {\it not} reproduce the perturbative results.

Let us now observe that the diagrams in Fig.~3 generate a new assisted
tunneling process ($ii$) originally absent from the starting
Hamiltonian Eq.~(\ref{eq:Hamilt}):
\begin{equation}
H_{\rm asst.}={1\over 2} \Delta_1 \tau^x (c_e^+ c_e - c_o^+
c_o) = {1\over 2} \Delta_1 \tau^x (c_1^+ c_2 + c_2^+
c_1)\;,
\label {eq:H_asst.}
\end{equation}
where the annihilation operators for the orthogonal left and right electron
states $c_{1,2}$  have been defined following
Ref.~\onlinecite{MF2} as
\begin{equation}
c_{1,2} =(c_e \pm c_o)/ \sqrt{2}\;.
\label{eq:site}
\end{equation}
The contributions of the diagrams in Fig.~3 to the scaling equations
depend on whether the intermediate  electron line is even or odd and
can be calculated by rescaling the cutoff $D\rightarrow D- dD$
(see Ref.~\onlinecite{Vlad_Zaw})
\begin{equation}
\Delta_0 V_3^2 \left[ \int_{D-dD}^D - \int_{-D}^{-D+dD} \right]
{1\over \epsilon^2} \varrho_{e,o}(0)(1 + \alpha_{e,o} \epsilon)
d\epsilon \;,
\label{eq:generation}
\end{equation}
and generate the following terms
\begin{equation}
\sim  \Delta_0 V_3^2 \tau^x {dD \over D}\left \{
{\varrho_e\alpha_e +\varrho_o\alpha_o\over 2}(c_e^+ c_e + c_o^+ c_o)
+ {\varrho_e\alpha_e - \varrho_o\alpha_o\over 2} (c_e^+ c_e - c_o^+ c_o)
\right\} \; .
\label{eq:correction}
\end{equation}
 Here the first term is a dynamical renormalization of the tunneling
amplitude which commutes with the screening term $V_3$, but the
second term is a generated assisted tunneling.

Thus if $\varrho_e\alpha_e \not = \varrho_o \alpha_o$ then the two
diagrams in Fig.~3 generate the assisted tunneling process.
For $\varrho_e = \varrho_o$ the condition $\alpha_e \not =
\alpha_o$ is equivalent with the one that in the site
representation $c_{1,2}$ the Green's function for the
conduction electrons has also {\it offdiagonal} matrix elements.
This result is in accordance with the suggestion of Moustakas
and Fisher who related the appearance of the assisted tunneling
to cross scattering between the electron channels '1' and '2' and
in this way to
the non zero value of the coupling $V_2$ (see
Eq.~(\ref{eq:spfunc})).\cite{MF1} However, as one can see from
Eq.~(\ref{eq:correction}) one has a logarithmic contribution
only if the {\it electron-hole} symmetry is broken in the even
and odd channels in different ways, which means that this is a
combination of $1\to 2$ scattering with the e-h symmetry breaking
which generates the assisted tunneling. (In the simple model of
Ref.~\onlinecite{MF1} the term $V_2$
generated {\it both} cross scattering between
channels 1 and 2 {\it and} the e-h symmetry breaking.) Another important
difference is that the term generated is proportional to $g^2$
in contrast to the scaling equations of Ref.~\onlinecite{MF1,MF2},
where such a coupling is also generated in the $g=0$ case, where
the conduction electron sector and the HP is dynamically
completely decoupled. This indicates that, although the scaling
equations of Ref.~\onlinecite{MF1} leave the free energy of the system
invariant, {\it the scaled couplings} there have {\it nothing to do with
physically measurable quantities}  like the scattering
amplitudes directly connected to the vertex function, and must
be interpreted with care.

Thus for $\alpha_e \ne \alpha_o$ the assisted tunneling term is
really generated and it makes the marginal fixed point
unstable.\cite{Zawa} However, its coefficient is very small as
$g\sim 0.1$ and $\Delta_0/ D_0 \sim 10^{-4}$, and therefore it
has always been dropped in the previous calculations, even if it
gives a logarithmic contribution (see
Refs.~\onlinecite{BVZ,Vlad_Zaw}).  It will be shown later that
this amplitude is too small to generate an assisted tunneling
which brings the TLS into the vicinity of the two-channel Kondo
fixed point described in the introduction when the hopping rate
$\Delta_0$ is also taken into account as a lower cutoff for the
scaling.

It is worth mentioning that the e-h symmetry can also be broken
in the case of a constant density of states but by using different upper and
lower cutoffs $D_{\rm upp}$ and $-D_{\rm low} $. In this case,
however, the contribution is not logarithmic.

\section{Analysis of the additional symmetry valid for
electron-hole symmetry in the absence of the electron assisted
hopping $\Delta_1$}
\label{sec:newsymmetry}

In the present section we analyze the symmetry properties of the
commutative TLS problem and show that the model has an
additional very restrictive symmetry related to the generation
of the assisted tunneling.  This symmetry is closely related to
the conservation of the number of the conduction electrons in
channels '1' and '2' introduced by Moustakas and Fisher.\cite{MF1,MF2}

The formulation of the e-h symmetry needs special care. In the
general case the e-h symmetry does not hold exactly. There is,
however, an approximate transformation  nearby the Fermi surface
which connects electron and hole states at the same part of the
Fermi surface. In this approximate transformation the density of
states is assumed to be independent of the energy and the
momentum dependence of the different phase factors and wave
functions is neglected, therefore it holds only in a very
restrictive sence. The transformation is illustrated in
Fig.~\ref{fig:symmetry}, where a small section of the Fermi surface at
the Fermi wave vector is considered. 

The deviations in the
normal directions at that small section are neglected and the
dispersion is linearized along the normal direction. 
Thus the electron state with momentum ${\bf k}$ and energy
$E({\bf k})$ can be characterized by a wave vector at the 
Fermi surface, ${\bf k}_F$ and the energy $\epsilon= E({\bf k})-E_F$:
\begin{equation}
{\bf k} = {\bf k}_F + {{\bf v}_F({\bf k}_F) \over
|{\bf v}_F({\bf k}_F)|^2} \epsilon \; ,
\end{equation}
where  ${\bf v}_F({\bf k}_F)$ is the Fermi velocity at
point ${\bf k}_F$. Then the e-h transformation can be
written as
\begin{equation}
c_{{\bf k}_F, \epsilon} \to c^+_{{\bf k}_F,-\epsilon} \;.
\end{equation}
In that approximation the energy independent density of states
can be written as
\begin{equation}
\varrho_0 = {1\over (2\pi)^3} \int {d S_{{\bf k}_F} \over |{\bf
v}_F({\bf k}_F)| } \;,
\end{equation}
where $dS_{{\bf k}_F} $ is the Fermi surface element. That
approximation becomes more accurate as the Fermi surface is
approached. Thus a systematic expansion around the Fermi surface
can be developed, where the zero order contribution is given
above and the further terms can be given in terms of $\epsilon$.
Such expansion has been used in the previous Section for the
local density of states by introducing the $\alpha_{e,o}$
factors. 

As an illustration the electron annihilation operators
at $\pm {\bf R}/2$ are considered for $|{\bf R k}_F|\ll 1$, thus
the phase factors are slowly varying functions of $\epsilon$.
Now $c_{\pm}$ is defined similarly to Eq.~(\ref{eq:c_pm}):
\begin{equation}
c_{\pm} = \int {d\epsilon \over 2\pi}\int {d S_{{\bf k}_F} \over
(2\pi)^2 } {1 \over v({\bf k}_F)} e^{ \pm i( {\bf k}_F +
\epsilon {\bf v} ({\bf k}_F)/ {\rm v}^2({\bf k}_F)) {\bf R}/2}
c_{{\bf k}_F,\epsilon} \; .
\end{equation}
Ignoring the $\epsilon$-dependence of the exponent the e-h
stransformation takes the simple form $c_{\pm} \to (c_{\mp})^+$,
which combined with Eqs.~(\ref{eq:c_e,o}) and (\ref{eq:site})
gives the simple form
\begin{equation}
c_1 \to c_2^+ \;.
\end{equation}
In the calculation of the infrared divergencies in the leading
order the approximate e-h symmetry can be applied. The
corrections in the  sense discussed above are less divergent or
even convergent due to the occurring extra powers of $\epsilon$.
Thus the above approach provides a systematic expansion in
breaking of the e-h symmetry for an arbitrary Fermi surface.

There are, however, special cases, where exact e-h
transformations exist, e.g., for a half filled cubic tight
binding model discussed in
Appendix~\ref{Noz}.\cite{Affl_Jones,Affl} These transformations
 differ essentially from the above-discussed one as different
regions of the Brillouin zone are connected by them.

In order to study the interaction first its structure must be discussed.
The following combinations of the creation and annihilation
operators of the HP span a complete Hilbert space, and thus are conveniently
described by Pauli matrices
\begin{eqnarray}
&&d^{\dag}_1 d_1 +d^{\dag}_2 d_2 \leftrightarrow {\rm I}=\tau^0 \;,\\&&
d^{\dag}_1 d_1 -d^{\dag}_2 d_2 \leftrightarrow \tau^z \;,\\&&
d^{\dag}_1 d_2 +d^{\dag}_2 d_1 \leftrightarrow \tau^x \;,\\&&
d^{\dag}_1 d_2 -d^{\dag}_2 d_1 \leftrightarrow i\tau^y \;,
\end{eqnarray}
where I is the unit matrix. A suitable choice of a basis for
the electron operator products $c^{\dag}_{i\sigma}c_{j\sigma}$
$(i,j=1,2)$ is:
\begin{eqnarray}
&&O_1=c^{\dag}_1 c_1 -c^{\dag}_2 c_2 =c^{\dag}_e c_o +c^{\dag}_o c_e \;,
\nonumber \\
&&O_2={\textstyle{1\over i}}\left(c^{\dag}_2 c_1 -c^{\dag}_1 c_2\right)
={\textstyle{1\over i}}\left(c^{\dag}_e c_o -c^{\dag}_o c_e\right) \;,\nonumber \\
&&E_1=c^{\dag}_1 c_1 +c^{\dag}_2 c_2 =c^{\dag}_e c_e +c^{\dag}_o c_o \;,
\nonumber \\
&&E_2=c^{\dag}_1 c_2 +c^{\dag}_2 c_1 =c^{\dag}_e c_e -c^{\dag}_o c_o \;,
\label{eq:operators}
\end{eqnarray}
where $E_i$ are symmetric and $O_i$ antisymmetric under the transformation
$1\leftrightarrow2$.

Now the following assumptions are made:\\
(i) the interaction between the atom and the electrons is invariant
under the  parity transformation between left and right.\\
The energy of the atom at left and right positions may be different
thus the tunneling atom may sit in an asymmetric potential well.
The assumption (i) is obvious for the screening interaction
(see the term proportional to $V_3$ in Eq.~(\ref{eq:int})) and the assisted
tunneling  exhibits the same invariance
(see Eq.~(\ref{eq:H_asst.})).\cite{Vlad_Zaw}
 This symmetry poses a constraint on the possible combinations of
operators in the Hamiltonian:
\begin{eqnarray}
&&H^z =\tau^z (V^z_1 O_1 +V^z_2 O_2) \;,\\
&&H^x =\tau^x (V^x_1 E_1 +V^x_2 E_2) \;,\\
&&H^y =\tau^y (V^y_1 O_1 +V^y_2 O_2) \;,
\end{eqnarray}
and the last combination, proportional to $\tau^0$ is not given since it
can be incorporated into the potential scattering.
The $V^\alpha_i$'s ($i=1,2$, $\alpha=x,y$) are the appropriate
couplings for assisted tunneling. The simplest form of the
screening interaction is described by $V^z_1$, where the
scattering on the atom is $s$-type.

We note at this point that the most general form of the Hamiltonian
may contain operators which can not be expressed simply by means of the
operators in Eq.~(\ref{eq:operators}) because the operators $c_1$ and $c_2$
 have simple specific structure in the momentum space
(see Eqs.~(\ref{eq:c_(e,o)}) and (\ref{eq:site})), and interactions
containing more elaborate momentum dependence can not be
described in terms of these operators. 

A specific commutative model can be defined by adding to the general
assumption (i) the followings\\
(ii) The electron hole symmetry holds for the kinetic energy part of
the Hamiltonian: $\varrho_e(\epsilon)=\varrho_e(-\epsilon)
=\varrho_o(\epsilon)=\varrho_o(-\epsilon)$.\\
(iii) \underline {Additional symmetry}: 
The interaction is also symmetric under the transformation:
$c_1\leftrightarrow c^{\dag}_2$, $c_2\leftrightarrow
c^{\dag}_1$. As discussed above this symmetry transformation
is connected to the approximate local e-h symmetry of the Fermi
surface. \\
(iv) \underline {Time reversal symmetry}, which connects the creation
operators of the time reversal states $\lambda$ and $\lambda^T$,
$a^{\dag}_\lambda\to a_{\lambda^T}^{\dag}$,
and replaces any C-number by its complex conjugate as $c\to c^\ast$.

Assuming that the wave functions of the tunneling atom are real
($\lambda^T=\lambda$)
\begin{equation}
d_1^T=d_1 \;,\; d_2^T=d_2 \;.
\end{equation}

Using the definition of time reversal symmetry for the free
electron operators $(c_{\bf k})^T=c_{-\bf k}$ and
Eq.~(\ref{eq:c_pm}) we find that $(c_{\pm k})^T=c_{\pm k}$, and
similarly $c^T_1=c_1$ and $c^T_2=c_2$. 
\hfill \\
(v) Moustakas and Fisher pointed out, that in the present case the
transformation $c_1\to e^{i\phi_1}c_1$ and $c_2\to e^{i\phi_2}c_2$
also leaves the Hamiltonian invariant. This symmetry is
connected to the conservation of the number of electrons in
channels 1 and 2. It actually follows from the previously
mentioned assumptions ($i-iv$).

For completeness we mention that in assumption (iii) the left-right
transformation for the electron could be replaced
by performing it in the atomic variable ($d_1\leftrightarrow d_2$)
instead of $c_1\leftrightarrow c_2$.
In that case, however, the potential well must be symmetric, thus
the atomic energies must be degenerate ($\Delta^z=0$).
For symmetry (iii) only the exchange of the relative positions of the
HP and the electrons are important.

As a consequence of the time reversal symmetry (iv) $V_2^z=0$,
$V_2^x=0$ and $V_1^y=0$.
The additional symmetry (iii) rules out all the terms except those
proportional to $O_1$, thus that excludes the assisted tunneling.
In that way the symmetries (iii) and (iv) allow only the interaction
term of coupling $V_1^z$, thus they ensure the commutativity of the model
even for an asymmetric TLS Hamiltonian ($\Delta^z\ne0$).

All the above consideration hold not only for the initial Hamiltonian,
but for all the renormalized vertices. {\it Thus the commutative model
remains commutative during the scaling as a consequence of symmetries}
(ii)--(iv), and no terms can be generated which are different from
$V_1^z$, and $\Delta_0$.  As also mentioned in the Introduction
the elimination scheme of Moustakas and Fisher {\it also}
violates this symmetry property of the Hamiltonian and generates
terms even for the commutative TLS with the maximum symmetry
(i)-(iv), which are not allowed.

\section{Instability of the marginal line due to electron-hole
symmetry breaking}
\label{sec:instab}

It has been argued in Sec.~II that the {\it marginal} line of fixed points
is unstable due to breaking of the e-h symmetry.
There is, however, an infrared cutoff appearing as the energy splitting
$\Delta_0$ between the symmetric and antisymmetric heavy particle states
as a consequence of the direct, spontaneous hopping between the left
and right states.
The asymmetry of the potential well $\Delta_z$ further increases
that cutoff to $\Delta=(\Delta_0^2+ \Delta_z^2)^{1/2}$.
Thus the same direct hopping $\Delta_0$ is responsible for the instability of
the fixed point and for blocking the scaling  at lower energies.
It will be shown, that for a TLS with realistic parameters in metals
the scaling is restricted to the neighborhood of the marginal line
which is far from the two channel fixed point. In order to demonstrate
this statement in this section the notation of the TLS literature is
used,\cite{Vlad_Zaw} which results in much more symmetrical scaling equations.
With these notations the  interaction part of the Hamiltonian is
\begin{equation}
H_{\rm int}=\sum_{i=0,x,y,z}\sum_{\alpha,\beta}
V^i c^{\dag}_{\alpha\sigma}\sigma^i_{\alpha\beta}c_{\beta\sigma}
\tau^i  \;,\label{41}
\end{equation}
where $\alpha,\beta=1,2$ label the orbital degrees of freedom and the spin
index $\sigma$ has been restored. (Note that the matrix
$\sigma^i$ is acting in the {\it orbital} indices of the
conduction electrons.)
The $V^i$'s denote the TLS-electron couplings, and $\sigma^0$ denotes
the unit matrix. Note that now  $V^0$ is initially set to zero,
since its effect is incorporated in the local electronic propagators.
Assuming $\varrho_e(0)=\varrho_o(0)=\varrho_0$ for the sake of
simplicity, the dimensionless
values of the couplings are defined as $v^i=\varrho_0 V^i$. This
approximation is not essential, but makes the equations more transparent.
These couplings of
Ref.~\onlinecite{Vlad_Zaw} can be expressed with the notations of
the previous sections as $V^z\to V^3$, $V^x\to\Delta_1$ and
$V^0\to V_1$, $V_2=0$.
In the e-h symmetric case $\alpha_{e,o}=0$ the scaling equations
of the next to leading logarithmic  order  are \cite{Vlad_Zaw,cikk_2}
\begin{eqnarray}
&&{dv^0\over dx}=0 \;,\label{42}\\&&
{dv^x\over dx}=4v^y v^z -8 v^x ((v^y)^2 + (v^z)^2) \;,\label{43}\\&&
{dv^y\over dx}=4v^z v^x -8 v^y ((v^x)^2 + (v^z)^2) \;,\label{44}\\&&
{dv^z\over dx}=4v^x v^y -8 v^z ((v^x)^2 + (v^y)^2) \;,\label{45}
\end{eqnarray}
where $x=\log D_0/D$ with the scaled band width $D$, and the
scaling of the splitting is described by
\begin{equation}
{d\Delta_0\over dx}=-8\Delta_0\bigl[(v^y)^2+(v^z)^2\bigr] \;.
\end{equation}
(In Ref.~\onlinecite{Vlad_Zaw} instead of notation $\Delta_0$ the
parameter $\Delta^x = \Delta_0$ is used.) Typical starting values
of these
parameters can be estimated as  $v^z\sim 0.2$, $v^x < 10^{-3}-10^{-4}$,
$v^y=0$, $\Delta_0\sim 1-10\,$K, $D\sim 10^5\,{\rm K}\sim 10\,$eV.

The breaking of e-h symmetry can be taken into account in two different ways:
(i) a slope in the density of states and (ii) different lower
and upper cutoff, $D_{up}$ and $D_{\rm down}$. (One can also
imagine a combination of the previous two cases).

First we consider the simplest case (i). In this case the general scaling
equations are quite complicated,\cite{cikk_2} however, in the limit
$|v^z|\gg|v^x|,~|v^y|$, they can be linearized in the 'small
couplings', $\Delta_0$, $v^x$ and $v^y$, and one can convince
himself easily that the discussed
diagrams only modify the linearized version of Eq.~(\ref{43}).
Then using the result given by Eq.~(\ref{eq:generation}) and
combining that with Eqs.~(\ref{42}--\ref{44}) one gets
\begin{eqnarray}
{d \over dx}{\Delta_0} &=&
 -8(v^z)^2 {\Delta_0} \;, \\
{dv^x\over dx} &=& 4v^y v^z - 8 v^x ({v^z})^2 +
\;\delta\alpha\;{\Delta_0} {v^z}^2 \;,\label{49} \\
{dv^y\over dx}&=&4v^x v^z -8 v^y ({v^z})^2 \;,
\end{eqnarray}
where $\delta\alpha=\alpha_e-\alpha_o$ and $v^z\approx const$.
These linear differential equations can be easily solved and one
obtains for not too large $v^z$'s:
\begin{eqnarray}
\Delta_0 (x) &=& \Delta_0
\left( {D_0 \over D }\right)^{-8 (v^z)^2} \;, \\
v_x(x ) & \approx & \delta\alpha \; {v^z \over 4}
\Delta_0  \left( {D_0 \over D }\right)^{4 v^z -8 (v^z)^2}
\;.
\end{eqnarray}
The scaling stops when $(\Delta_0/D)(x) \sim 1$. Using the
typical values given above and $\alpha\sim 1$ (See
App.~\ref{app:A}) the value of the generated coupling $v^x$ at
the freezing out of the TLS can be estimated as
\begin{equation}
v^x_{\rm final} \sim 10^{-3} - 10^{-4} \;,
\end{equation}
which is too small to give an observable effect.
This situation is sketched in Fig.~\ref{fig:scaling}.

In the last part of this section we briefly discuss case (ii) with
$\alpha_{e,o}=0$, $\varrho_e(0)=\varrho_o(0)$ but with
$D_{\rm up} \gg D_{\rm low}$ for both the even and odd channels,
which means a very asymmetrical band. In this case a
two-cutoff scaling can be used. The derivation of the scaling
equations can easily generalized for the case of this
asymmetrical cutoff. Assuming $D_{\rm up}>D_{\rm low}$ the
integration now has to be carried out first for the energy range
$D_{\rm low}<|\omega|<D_{\rm up}$ and the couplings obtained by
scaling must be then used for a symmetric model with
$D=D_{\rm low}$ as starting values. Now the results of Sec.~II can
be generalized by keeping only the first part of
integral (\ref{eq:generation}).
Thus the coupling $v^x$ at $D=D_{\rm low}$ ($D_{\rm up}>D_{\rm low}$) is
\begin{equation}
V^x(D=D_{\rm low})\sim \Delta_0(0)(V^z(0))^2
\Biggl({1\over D_{\rm low}}-{1\over D_{\rm up}}\Biggr)
\left( {\varrho_e(0)- \varrho_o(0)} \right) \;.
\end{equation}
As for a realistic metallic case $D_{\rm up}\sim D_{\rm low}\sim
10 \;{\rm eV}$ and
$\varrho_e(0) \sim \varrho_o(0) \sim 1/D $ the corresponding
dimensionless assisted tunneling can be estimated
as $v^x(D=D_{\rm low})\sim 10^{-6}$.
That initial value is too small to be increased essentially by scaling
to obtain a value at the lower cutoff the scaling $D=\Delta_0$
which is comparable with the two channel fixed point $v^x\approx 0.25$~.
Indeed using Eq.~(2.17b) of Ref.~\onlinecite{Vlad_Zaw} one obtains
\begin{equation}
v^x(D=\Delta_0)=v^x(D_{\rm low})\cosh(4v^z x) \;,
\end{equation}
where $x=\log(D_{\rm low}/\Delta_0) \sim10$, thus
\begin{equation}
v^x(D=\Delta_0)\sim 6\cdot10^{-4} \;,
\label{eq:est}
\end{equation}
which is still smaller by a factor $\sim 10^{-3}$ then the fixed point
value $v^x\sim0.25$~.
Finally it should be mentioned that the spontaneous tunneling rate
$\Delta_0$ also renormalizes downward, but not more then 1--2 orders of
magnitude, and  since the typical renormalized value of $\Delta_0$
is $\sim 1 {\rm K}$ the estimation (\ref{eq:est}) is not
essentially modified by the renormalization of $\Delta_0$.

In the discussion above it was assumed that $D_{\rm up}/D_{\rm low}$
can not be very large.
That is not the case of a degenerate semiconductor, as a Pb$_{1-x}$Ge$_x$Te
crystal with $x\ll 1$, where orbital Kondo effect has been
observed.\cite{Fukuyama}
In such a case even $D_{\rm up}/D_{\rm low}\sim100$ can be realistic and also
$\alpha$ can be present.
In that case an essential strength of the assisted tunneling can be
developed by scaling, even if it is negligible in the unrenormalized
Hamiltonian. That problem deserves further studies.

\section{Conclusions and discussion of the path integral technique}
\label{sec:conclusions}

In the previous sections the instability of the commutative TLS model
against the breaking of the e-h symmetry has been studied using
the multiplicative renormalization group.  We found that without this
symmetry  breaking the line $V_x=V_y=0$ is marginally stable.

The instability of this marginal line has been shown by Moustakas and
Fisher,\cite{MF1} who  pointed out that the original commutative TLS
model has an artificial internal symmetry, connected to the conservation
of the electron numbers in the left and right channels defined
in Eq.~(\ref{eq:site}).
In the present paper we have shown that the model possesses an even more
restrictive additional symmetry reflecting the role of e-h
symmetry. This symmetry ensures the stability of the
previously mentioned marginal line in the presence of
e-h symmetry and in the absence of assisted tunneling.
The breaking of the e-h symmetry combined with the screening
$V_z$ and the spontaneous tunneling $\Delta_0$ leads to the
generation of assisted tunneling which drives the system 
in the direction of 
the two-channel Kondo fixed point where the two channels are due
to the double degeneracy of the conduction electrons because of
their real spin.  However, as it has been shown in
Sec.~\ref{sec:instab}, this fixed point can never be reached by
the present mechanism as the driving force for the instability
and the energy scale stopping the scaling are both provided by
the spontaneous tunneling, $\Delta_0$.

It has also been pointed out that there is an ambiguity in the
definition of the e-h symmetry transformation. Our symmetry ($iii$) is
valid for the kinetic part of the Hamiltonian in an arbitrary
system nearby the Fermi surface if the curvature of the Fermi
surface and the momentum dependence of the conduction electrons'
energy are neglected, i.e., a local e-h symmetry is assumed at
the Fermi surface, and the two minima of the TLS are close to
each other: $Rk_f\ll 1$.  As pointed out, this kind of
transformation can also be realized by an exact e-h symmetry
transformation of a half-filled tight binding model.\cite{Affl}

We have also estimated how far the system can get from the
marginal line. For a conventional metal the dimensionless
measure of the e-h symmetry breaking term can be estimated
to be of the order of one, $\alpha\sim 1$. Then the generated
assisted tunneling at the freezing out of the TLS motion is of
the order of  $v_x\sim v_y\sim 10^{-3}-10^{-4}$, which are too
small to produce an observable effect and the system is still
not displaying the two-channel Kondo behavior. On the other
hand, for degenerate semiconductors the high asymmetry in the
lower and upper cutoffs may generate a much larger $v_x$ via
$\Delta_0$, which might be responsible for the Kondo effect
occurring in such materials.\cite{Fukuyama}

The above discussion leads to the conclusion that
for a TLS in an ordinary metal a strong enough
initial assisted tunneling is required in addition to
the Hamiltonian (\ref{eq:Hamilt}) to get a scaling into the
intermediate vicinity of the two-channel Kondo fixed point and
to have logarithmic minima in the electrical
resistivity.\cite{Vlad_Zaw} The required assisted
tunneling can also be generated via virtual
excitations to the excited states of the HP.\cite{ZarZaw} The latter
mechanism seems to be very promising, however, further studies
are needed to check that these virtual hoppings do not generate
a too large splitting $\Delta_0$ which might block the formation
of the resistivity minimum.

We stress at this point that the role of $\Delta_0$ is
completely different in the commutative model and the
non-commutative model with $V_x,V_y \not= 0$.\cite{Vlad_Zaw} In the
latter case $\Delta_0$ plays basically the role of an infrared cutoff parameter in
the scaling and it drives the system from a
non-fermi-liquid to a fermi-liquid transition.\cite{Vlad_Zaw} This
crossover has also been observed in recent point contact
measurements, where it has also been possible to tune the crossover
parameter $\sim \Delta_0$ by electron
migration.\cite{Buhrmanlast}  This cross-over, of course, is
only observable if $\Delta_0$ is not too large to stop the
evolution of the non-Fermi liquid properties.
We also mention that in the very strong coupling region
$\Delta_0$ might become irrelevant, thereby playing again a
qualitatively different role. This situation, however, is very
unlikely to occur for realistic model parameters.

The present calculations support the general considerations of
Moustakas and Fisher\cite{MF1,MF2} concerning the instability of the marginal line,
there are, however, essential differences. The source of these
differences can be due to the ambiguities in the scaling
procedures based on the path integral method.\cite{VladZimZaw,MF1,MF2}
Since this technique is well established in the literature,
we only briefly sketch the procedure used.

In the path integral method \cite{VladZimZaw} the
partition function  of the system is written in an imaginary time path
integral form,  and is factorized as $Z= Z_1 Z_2$,  where $Z_1$
involves only the screening operator $V^z$,  while all the
hopping terms are pushed into the second term,  $Z_2$,  which is
calculated perturbatively in the hoppings. Following the same
lines as Nozi\`eres and de Dominicis in the solution of the
X-ray absorbtion edge problem  \cite{NozDeDom}
one can integrate over the electronic degrees of freedom and
obtain an effective functional for the TLS path, containing
logarithmic interactions between the (spontaneous and assisted)
flips of the TLS.\cite{VladZimZaw}
Finally, one can generate scaling equations for the `fugacities'
(corresponding to the different couplings and hopping
amplitudes) by reducing the high energy cutoff intercoming in
the conduction electrons Green's  function.\cite{NozDeDom,YuvAnd}

The ambiguity of the path integral technique lies in the last
step of the procedure, i.e. in the elimination of the high
energy degrees of freedom, which corresponds to the change of a
small time scale $\tau_0 \sim 1/\epsilon_F$ in the path integral scheme.

In general two different elimination procedures can be applied:

(i) One eliminates events of any kind which are closer in the
imaginary time than the cutoff $\tau_0$.\cite{MF1,MF2}

(ii) One eliminates only those events which are closer to each
other than $\tau_0$ {\it and} in which electrons are
involved.\cite{VladZimZaw}

Method (i) is followed by Moustakas and Fisher\cite{MF1,MF2} and in many other
applications. It has been applied, e.g., for the problem of 1D
disordered metals\cite{GiamSchultz} where a special scheme had
to be applied to avoid artificial generation of
interactions. The aforementioned case of Ref.~\onlinecite{MF1} is
somewhat similar. On the other hand method (ii) rather
corresponds to Anderson's poor man scaling, where the electronic
heat bath is eliminated step by step via the reduction of the
bandwidth cutoff $D$.\cite{Anderson}

To show that method (i) delivers unphysical results let us
consider the path of the TLS in Fig.~\ref{fig:path} where a potential
scattering occurs immediately after a spontaneous tunneling.
Using method (i) Moustakas and Fisher replaced the event inside
the box by a generated assisted tunneling interaction (see also
Fig.~2 in Ref.~\onlinecite{MF1}). The artificial nature of this
assisted tunneling is obvious since it is generated
even if the dynamics of the TLS and the electrons are completely
decoupled ($v_x=v_y=v_z=0$).
This generated interaction, of course, can not be related to any
measurable dynamical process. On the other hand, method (ii)
does not generate this artificial interaction.

The path integral method and the diagrammatic renormalization
group are essentially different with respect to dynamical
quantities. In the path integral approach, in principle,
both methods (i) and (ii) can be followed, to calculate {\it
thermodynamical quantities} even if method (i) is more
frequently used.  As a matter of fact in Appendix B we show that
the scaling equations for the two-electron scattering
obtained by methods (i) and (ii) are related by a formal
transformation in the small coupling limit. The relationship between the different dynamical
quantities is, however, much more complicated. To see this it is
enough to consider 
the HP Green's function $D(\tau)=\langle  d^+_1(\tau)d_2(\tau)
d_2^+(0)d_1(0)\rangle$. While this Green function is invariant under
the transformation (ii), using method (i) it acquires
corrections of the type  $\delta D \sim \langle  d^+_1(\tau)d_2(\tau)c^+_i(\tau)
c_j(\tau) d_2^+(0)d_1(0)\rangle$ and  $\sim \langle  d^+_1(\tau)d_2(\tau)c^+_i(0)
c_j(0) d_2^+(0)d_1(0)\rangle$ already in the first step of the
renormalization group ($i,j= 1,2$), and under the subsequent
steps it becomes a very complicated object composed from
different many particle propagators.

The multiplicative renormalization groop, on the other hand,
{\it guarantees} the invariance of the different Green's
functions. Thus the best way to choose the correct elimination
procedure in the path integral technique is to compare it to the
perturbative results. This comparison gives also the result that the {\it it is
method (ii) which gives the correct dynamical quantities}.
Therefore the scaling equations derived in
Refs.~\onlinecite{MF1} and \onlinecite{MF2} must be interpreted
with care.

This conclusion may serve as a hint for other models where the
strength of dynamical processes must be determined.

Finally, we shortly discuss the case of two-electron scattering
$\Delta_2$. It has been raised in Refs.~\onlinecite{MF1,MF2}
that this process becomes relevant at the two-channel Kondo
fixed point. While this statement seems to be correct in the
highly anisotropical limit it is questionable in the strong
coupling limit, where both the Bosonization approach (used in
these works) and the Anderson-Yuval type scaling analysis 
loose their validity. On the other hand, this process has a
small amplitude initially and is highly irrelevant in the weak
coupling limit. This should be contrasted to the splitting of
the TLS which is relevant in {\it both} the strong and weak
coupling limits. Therefore, even if the two-electron scattering
would be a relevant process at the Kondo fixed point, the
dominant process driving away the TLS from the non-Fermi liquid
fixed point will be the splitting for a realistic TLS, usually
discussed in the literature. 

The generation of the $\Delta_2$ interaction term by the
e-h symmetry breaking can be analysed very similarly to that of the
assisted tunneling process. Our investigations show
that the generated $\Delta_2$ is of the order of
$v_z^4 \alpha^2 \Delta_0$ and in the region $v_z \le 0.4$
the two-electron scattering is frozen out before it aquires an
observable amplitude. On the other hand the role of $\Delta_2$
in the strong  screening case is still not completely clear
and deserves further studies.

To summarize, the general conclusion of the present study is,
that the breaking
of the e-h symmetry may have important consequences in cases,
where the e-h symmetry ensures the cancellation of some diagrams
(validity of Ward identities) and thus the
stability of some fixed points. In this way, the e-h symmetry
breaking may change the universality class of the problem, as it
actually does in case of the two-impurity Kondo
problem\cite{Jones_Varma,Affl_Jones,Silva,Kunose} and in
some other realizations of the two-channel Kondo problem.\cite{Affl_Jones,Affl}

\section{ACKNOWLEDGEMENT}
 G. Z. and A. Z. would like to
acknowledge D.L. Cox, F. Guinea, A. Moustakas, and A.W.W. Ludwig for
helpful discussions, and they are also grateful for the 
hospitality of the Institut Laue-Langevin, Grenoble, where
a part of the present work has been done. G. Z. has been
supported by the Magyary Zolt\'an Fellowship of the Hungarian
Ministry of Education. This research has been supported by the
grants No. OTKA T-021228, OTKA F016604 and J.F. No. 587.  G.T.Z. and
K.V. have been supported by the grants NSF DMR 95-28535 and OTKA
T-017128.

\appendix
\section{Calculation of the parameters $\alpha_{e,o}$}
\label{app:A}

In the present Appendix we show that the e-h asymmetry parameters
$\alpha_e$ and $\alpha_o$ are generated even in the absence of
potential scattering due to band structure effects.
A simple free-electron like parabolic band with mass
$m$ is assumed for the electrons.  The $V_3$-term of the Hamiltonian
can be derived by assuming a local electron-TLS interaction and it can be
written as
\begin{equation}
H_3 = V \left(\varrho({\bf R}/2) - \varrho(-{\bf R}/2) \right) \tau^z \; ,
\label{eq:app1}
\end{equation}
where $\varrho({\bf r})=\psi^+ ({\bf r})\psi({\bf r})$
denotes  the density of the conduction electrons
and we assumed that the average potential set by the TLS for the
conduction electrons has already  been taken into account in the calculation
of the parabolic electronic dispersion.
As the field operators are taken at the atomic sites a point-like
$s$-wave TLS-electron interaction is assumed. In the conventional way an upper
cutoff can be introduced by using a form factor in the momentum
representation which also makes the interaction non-local. The interaction part
(\ref{eq:app1}) can be rewritten by using the notations
$\psi_{e(o)}={1\over 2}\left( \psi\left( {\bf R} / 2 \right) \pm
\psi\left( -{\bf R} / 2 \right) \right)$ as
\begin{equation}
H_3 = 2V (\psi_e^{\dag} \psi_o + \psi_o^{\dag} \psi_e ) \; ,
\end{equation}
where the field operators can be written in a  momentum representation as
\begin{equation}
\psi_{e(o)} ={1\over (2\pi)^{3}} {1\over 2}
\int d^3{\bf k} \left( e^{i{\bf k R}/2} \pm e^{-i{\bf k R}/2}\right )
c_{\bf k}\; .
\end{equation}
 Note that these operators are different from the $c_{e,o}$ operators
defined in Eq.~(\ref{eq:c_(e,o)})
Then the spectral function for the Green's function $G_{ab}=
- \langle T\{\psi^{\dag}_a \psi_b\}\rangle = \delta_{a,b}G_a$, ($a,b= e,o$)
can be expressed as
\begin{equation}
\varrho_{e,o}(\omega) = {1\over (2\pi)^3} \int d^3{\bf k}
\delta(\omega -{{\bf k}^2 \over 2m})
\left\{
\begin{array}{c} \cos ^2({\bf kR/}2) \\
  \sin ^2({\bf kR/}2) \end{array} \right. \;.
\end{equation}
These integrals can be easily evaluated by taking the $z$-axis parallel to
${\bf R}$, and one finally obtains
\begin{equation}
\varrho_{e(o)}(\omega)= {\varrho(\omega)\over 2} \left[
1 \pm { \sin R (2m\omega)^{1/2} \over R (2m\omega)^{1/2}}\right]
\;, \label{eq:ro_e,o}
\end{equation}
where $\varrho(\omega) = m^{3/2} 2^{-1/2} (2\pi)^{-2} \omega^{1/2}$ is the
density of states for the free electrons for one spin direction,
and the factor in the bracket is related to the normalization factor
$N_{e,o}(k)$ given by Eq.(\ref{eq:normalization}). In
Ref.~\onlinecite{MF1} these factors taken at the Fermi surface
are incorporated in the
definition of the coupling $V_3$ as $(\varrho_e(\epsilon_F)
\varrho_o(\epsilon_F))^{1/2} \sim (1-(\sin k_F R\;/\; k_F
R)^2)^{1/2}$. There complete k-dependence can only be
taken into account by a k-dependent coupling breaking the e-h symmetry.

 Finally, from expression (\ref{eq:ro_e,o})
the $\alpha$ parameters at the Fermi energy can easily be
calculated as
\begin{equation}
\alpha_{e,o}= \left({ d \ln \varrho_{e,o} \over d
\omega}\right)_{\epsilon_F} = {1\over 2}{1 \pm \cos k_F R \over
1 \pm \sin k_F R\; / \; k_F R}\;\epsilon_F^{-1},
\end{equation}
with $k_F = (2m \epsilon_F)^{1/2}$. The parameters $\alpha_{e,o}$
are of the order of $\epsilon_F^{-1}$ and, e.g., for $k_F R \ll 1$
one obtains $\alpha_e=\alpha_o/3 = 1/2 \epsilon_F^{-1}$.

\section{Connection between the two scaling equations obtained
by the two different path integral methods}
\label{app:B}

In this Appendix we show that there exists a simple formal
correspondence between the small coupling limit of the scaling equations of
Refs.~\onlinecite{MF1} and \onlinecite{MF2} and those
 obtained by Vlad\'ar, Zim\'anyi  and Zawadowski\cite{VladZimZaw}.
This connection is based on the
artificial mixing of the TLS and conduction electron degrees of
freedom.

Throughout this Appendix, for the sake of compactness, following
Ref.~\onlinecite{Vlad_Zaw} we write the Hamiltonian in the
following form:
\begin{eqnarray}
H_{\rm int} &=&  \sum_{\mu,\sigma,\alpha,\beta}
c^+_{\alpha\sigma} V^\mu_{\alpha \beta} c_{\beta\sigma} \tau^\mu \;, \\
H_{\rm TLS} &=& \sum_{i=x,y,z} \Delta^i \tau^i \;,
\end{eqnarray}
where $\mu = 0,x,y,z$ and the indices $\alpha,\beta=1,2$ refer to the orbital
degrees of freedom of the conduction electrons. In the special
case considered in Sec.~\ref{sec:instab} the couplings take the
simple form $V^\mu_{\alpha\beta}=
V^\mu\cdot \sigma^\mu_{\alpha\beta}$. For the conduction
electrons a constant density of states $\varrho_0$ is used for both channels
$\alpha=1,2$.

Introducing the matrix notations $\varrho_0 V^\mu_{\alpha\beta}
\to {\underline v}^\mu$ the leading logarithmic scaling equations obtained by the
multiplicative renormalization group are\cite{Vlad_Zaw}
\begin{eqnarray}
{d{\underline v}_0\over dx} &=& 0 \nonumber \;, \\
{d{\underline v}_i\over dx} &=& 2 i \epsilon^{ijk}{\underline v}_j{\underline v}_k
\;,
\label{eq:genscale}
\end{eqnarray}
with $\epsilon^{ijk}$ denoting the Levi-Civita symbol. These
equations coincide with the small coupling limit of the scaling
equations of Ref.~\onlinecite{VladZimZaw} obtained by using
elimination scheme (ii) in Sec.~V.

The new couplings corresponding to Refs.~\onlinecite{MF1}
and \onlinecite{MF2} can be expressed as
\begin{equation}
{\tilde{\underline v}}^\mu \tau^\mu =
\left(1 - {\Delta^i \over D} \tau^i \right)^{-1/2}
{\underline v}^\nu \tau^\nu
\left(1 - {\Delta^i \over D} \tau^i \right)^{-1/2}\; ,
\label{eq:transf}
\end{equation}
where a summation must be carried out over repeated indices.
This transformation can be motivated as follows. In the
high-energy region $\omega\approx D$ the unperturbed
'dimensionless' TLS propagator\cite{Solyom}
$g^{(0)}(\omega) = \omega {\cal G}^{(0)} (\omega)$ can be
written as $g^{(0)}(D)\approx (1-\Delta^i\tau^i/D)^{-1}$
Transformation (\ref{eq:transf}) corresponds to mixing this
factor (which is part of the TLS Green's function) into the
vertex in an artificial way.

Writing out Eq.~(\ref{eq:transf}) explicitly and plugging into
Eq.~(\ref{eq:genscale}) we obtain the following scaling
equations for the ${\tilde{\underline v}}^\mu$'s
\begin{eqnarray}
{d{\tilde{\underline v}}_0\over dx} &=&  {\tilde{\underline v}}^i {\Delta^i \over D} \nonumber \;, \\
{d{\tilde{\underline v}}_i\over dx} &=& 2 i \epsilon^{ijk}{\tilde{\underline
v}}_j{\tilde{\underline v}}_k + {\tilde{\underline v}}^0 {\Delta^i \over D}
\;.
\label{eq:tildascale}
\end{eqnarray}
Replacing the simple couplings of Sec.~\ref{sec:instab} into
these equations and putting $\Delta^y=\Delta^z=0$ one obtains
exactly the expanded version of the scaling equations in
Ref.~\onlinecite{MF2}. We note at this point as well that terms
proportional to $\sim \Delta^j/ D$ give no contribution to the
universal properties of
the the model and can not be taken into account by a simple
renormalization group procedure since they vanish in the scaling
limit $D\to \infty$, ${\underline v^\mu}(D_{\rm ref}), \Delta^i(D_{\rm
ref})=cst$, $D_{\rm ref}$ being a reference energy scale kept
constant (it can be chosen to be the Kondo temperature, e.g., in
the non-commutative model). One can also show easily that the
solution of the scaling equations derived from terms like $\sim
{1\over D} \ln {D\over \omega}$ does not reproduces the
perturbative results and is inconsistent with the scaling
hypothesis. Therefore the terms generated in
Eq.~(\ref{eq:tildascale}) by the transformation should vanish in
the scaling limit.

\section{Mixing of the high- end low-energy states in a simple
potential scattering model}

In this appendix we show how the original low- and high-energy
degrees of freedom are mixed up by a simple potential scattering
and how the elimination procedure during the renormalization
group procedure should be carried out.  For this purpose we
investigate the simplest model possible: the scattering of
spinless Fermions by a Dirac delta scatterer. In this case only s-wave
scattering must be considered and the Hamiltonian can be written
as
\begin{eqnarray}
H & = & H_0 + H_U \;, \nonumber \\
H_0 &=& \sum_n \epsilon_n c_n^{+} c_n \;, \label{eq:ham} \\
H_U &=& {U\over L} \sum_{n,m} c^{+}_n c_m \;,\nonumber
\end{eqnarray}
where we assume for the sake of simplicity that the spectrum of
the unperturbed Hamiltonian $H_0$ is given by $\epsilon_n = {2\pi
\over L}(n + 1/2)$, $\{ n= -N,..., N\}$, $L$ being the linear size
of the system, and $c_n^{+}$ denotes the creation operator of
a conduction electron with label $n$. This assumption
corresponds to a constant density
of states (DOS) between the high- and low-energy cutoffs $\pm D
= \pm {2\pi N / L}$.

Eq.~(\ref{eq:ham}) can be diagonalized easily via the unitary
transformation
\begin{equation}
c_\epsilon = \sum_n \gamma(\epsilon,n) \; c_n \; ,
\end{equation}
where the operator $c_\epsilon$ annihilates an electron with
energy $\epsilon$ and $\sum_n \gamma(\epsilon,n)
\gamma(\epsilon^\prime,n)=\delta_{\epsilon,\epsilon^\prime}$.
The energy spectrum is determined by the self-consistent
equation
\begin{equation}
1= {U\over L} \sum_n {1\over \epsilon -\epsilon_n} \;.
\label{eq:spectrum}
\end{equation}
Then the factors $\gamma$ can be expressed as
\begin{equation}
\gamma(\epsilon,n) = {U \over L} {S(\epsilon) \over \epsilon
-\epsilon_n} \; ,
\label{eq:gamma}
\end{equation}
where the constant $S(\epsilon)$ is determined by the unitarity
condition
\begin{equation}
1=\sum_n \gamma^2(\epsilon,n) = {U^2 S^2(\epsilon) \over L^2}
\sum_n{1\over (\epsilon-\epsilon_n)^2}\;.
\label{eq:S}
\end{equation}
Eqs.~(\ref{eq:spectrum}), (\ref{eq:gamma}), and (\ref{eq:S})
constitute the complete solution of the problem.

Our main purpose is to determine the exact eigenstates and the
local density of states (LDOS) for the Hamiltonian above, which
is given by the spectral function of the Green's function
\begin{equation}
G(\tau)= - \langle T c(0,\tau) c^+(0,0)\rangle\;,
\end{equation}
where $c(0) = (\sum_n c_n)/\sqrt{L}$. This can be easily
expressed in terms of the parameters $\gamma(\epsilon,n)$ as
\begin{equation}
\varrho(\omega) = \sum_\epsilon \delta(\omega -\epsilon)
\left(\sum_n{1\over \sqrt{ L}}\gamma(\epsilon,n) \right)^2\;.
\label{eq:LDOS}
\end{equation}
The last factor in this expression is nothing but the squared of
the wave function amplitude $A_\epsilon= |\psi_\epsilon(0)|^2$
of the state with energy $\epsilon$.

As one can convince himself easily from the investigation of
Eq.~(\ref{eq:spectrum}) the spectrum of the total Hamiltonian
consists of $2N$ nearly equally spaced states forming a
continuum with energies $-D<\epsilon <D$ and a single bound
state\cite{Wolf} with energy
\begin{equation}
\epsilon_B = \pm D \pm {2 D\over e^{\pm 1/\varrho_0 U} -1}\;,
\end{equation}
where $\varrho_0 =1/2\pi$ and the '$\pm$' sign refers to
attractive and repulsive interactions, respectively. This bound
state gives a finite contribution to $\varrho(\omega)$ with a
weight
\begin{equation}
|\psi_B(0)|^2 = {\pi D \over U^2} {1\over {\rm sh}^2(\pi/U)}\; .
\end{equation}
While the energy spectrum of the continuum is hardly affected
by the potential scattering (see Eq.~(\ref{eq:spectrum})), the
wave function of the states at the impurity site is, and
therefore the continuous part of the LDOS becomes strongly
energy-dependent.  The selfconsistency equations for the states
in the continuum can also be solved exactly by using the
identity
\begin{equation}
\sum_{n= -\infty}^\infty {1 \over \epsilon -\epsilon_n} =
- {L \over 2} {\rm tg}\left( \epsilon L\over 2 \right)\;.
\end{equation}
Then the self-consistency equation (\ref{eq:spectrum})
simplifies to
\begin{equation}
1 = {U\over 2\pi}\left\{ \ln \left( D +\epsilon \over D
-\epsilon \right) - \pi {\rm tg} \left(\epsilon L \over 2 \right)
\right\} \;.
\end{equation}
In the limit $L,N \to \infty$, $D=cst$ the continuous part of
the spectrum is given by
\begin{eqnarray}
\epsilon & =& {2\pi \over L} (n+1/2) + {2\over L} \delta_n \;, \\
{\rm ctg} \delta_n & =& {2\over U} - {1\over \pi} \ln {D + \epsilon
\over D- \epsilon}\;,
\end{eqnarray}
and the coefficients $\gamma(\epsilon,n)$ can also be determined
exactly:
\begin{equation}
\gamma(\epsilon,n) = {2 \over L \; f(\epsilon)} {1\over \epsilon
- \epsilon_n }\; ,
\end{equation}
with
\begin{equation}
f(\epsilon) = \left[ \left( {1\over \pi}\ln{D +\epsilon \over D
- \epsilon} - {2\over U} \right)^2 + 1\right]^{1/2} \; .
\end{equation}

The wave function amplitude of the state with energy $\epsilon$
can also be determined in this limit:
\begin{equation}
A_\epsilon =|\psi_\epsilon(0)|^2 = {1 \over L}
\left[ \left( {U \over 2\pi}\ln{D +\epsilon \over D
- \epsilon} - 1 \right)^2 + {U^2 \over 4} \right]^{-1}
\;.
\label{eq:A}
\end{equation}
Replacing these amplitudes into Eq.~(\ref{eq:LDOS}) we reproduce
the results of the Green's function calculations
Eq.~(\ref{eq:spfunc}) and one can easily calculate the parameter
$\alpha$ as well:
\begin{equation}
\alpha={2U\over \pi (1 + U^2 / 4)} {1 \over D}\;.
\label{eq:alpha_pot}
\end{equation}

The calculation above reveals the origin of the e-h symmetry
breaking in the local properties. Eqs.~(\ref{eq:gamma}) and
(\ref{eq:A}) tell us that the original high-energy states of
$H_0$ are mixed into the true low-energy excitations of the
model with potential scattering.  If one tries to do the
renormalization group transformation by eliminate the
high-energy states of $H_0$ ($D\to D^\prime$) one immediately
deforms the wave function of the true low-energy excitations
and changes their coupling to the TLS drastically.
This manifests in the change of their amplitude at the origin (\ref{eq:A})
and the cutoff dependence of the parameter $\alpha$
(\ref{eq:alpha_pot}). Therefore the correct procedure is first to
diagonalize the total Hamiltonian $H_0 + H_U$ and then eliminate
the high-energy states of the  model.

\section{The e-h symmetry on a cubic tight binding model}
\label{Noz}

The half-filled cubic tight binding band shows the e-h
symmetry $\epsilon_{\bf k} = -\epsilon_{{\bf k} + {\bf K}}$ with
${\bf K} = (\pi/a, \pi/a, \pi/a)$. In this cubic case the
lattice can be divided into two sublattices $A$ and $B$. The
e-h transformation can be given in the site representation of
the electrons' creation and annihilation operators as $a_i \to
a_i^+$ and $b_j \to - b^+_j$ ($i\epsilon A$, $j\epsilon B$).
If one confines artificially  the positions of the TLS onto the
lattice sites even the non-commutative model is invariant under
the above e-h transformation combined with a left-right exchange
of the electronic onsite operators similarly to 
Ref.~\onlinecite{MF1}. This situation is similar to that of the
two-impurity Kondo model.\cite{Affl_Jones,Affl}

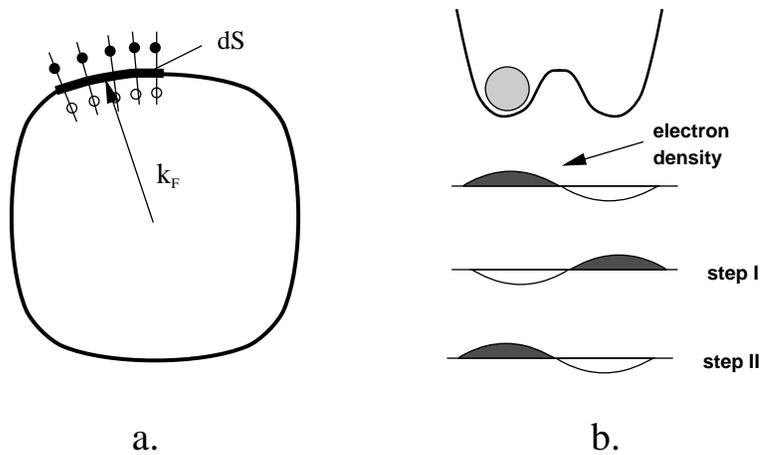
\begin{figure}
\epsfxsize=7cm
\hskip0.5cm
\vskip0.7truecm
\caption{(a) The electron and hole states (filled and open circles, 
respectively) connected by the approximate e-h transformation through the 
Fermi surface element $dS$. (b)
Schematic of the additional symmetry of the standard
TLS model consisting of an e-h transformation and a reflection
through the center of the TLS in the electronic degrees of freedom.}
\label{fig:symmetry}
\end{figure}

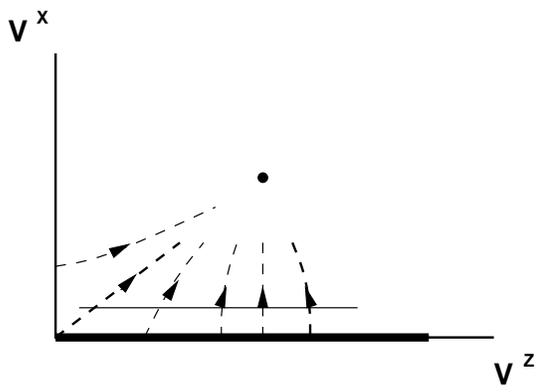
\begin{figure}
\epsfxsize=7cm
\hskip0.5cm
\vskip0.7truecm
\caption{Sketch of the scaling trajectories of the TLS. The
appropriate e-h symmetry breaking drives the TLS away from the
marginally stable fixed line $V_x=0$ ($\Delta_1=0$) towards the
two-channel Kondo fixed point. The scaling is stopped
by the renormalized splitting, and the final ground state is a
Fermi liquid. The freezing out of the TLS is indicated by a
light continuous line.}
\label{fig:scaling}
\end{figure}

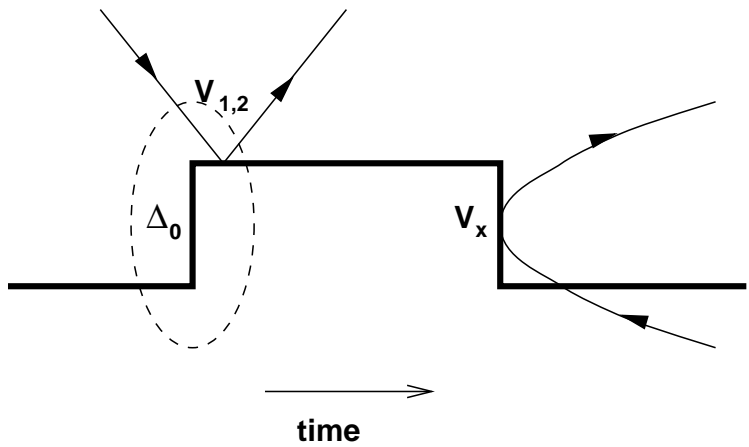
\begin{figure}
\epsfxsize=7cm
\hskip0.5cm
\vskip0.7truecm
\caption{The diagram resulting in the generation of artificial
interaction terms in the elimination scheme (i). The heavy line
represents the motion of the heavy fermion. The artificial
interaction is generated by the elimination of a spontaneous
tunneling event ($\Delta_0$) and a potential scattering close to
it ($V_{1,2}$). The lines with arrows represent the conduction
electrons.} \label{fig:path}
\end{figure}

The situation is drastically changed if
the tunneling atomic sites are not at the lattice points. For
the sake of simplicity we assume that the the TLS sites $\pm 
{\bf R}/2$ are between two neighboring lattice points labeled by
$(\pm a/2,0,0)$ and the corresponding operators are denoted by 
$a^+$ and $b^+$, respectively. In this case the electronic field at
$\pm {\bf R}/2$ is a combination of the atomic orbitals at sites
corresponding to $\pm a/2$. Assuming that the atomic orbitals
$\psi$ are s-tyme the field operators at sites $\pm {\bf R}/2$
are 
\begin{equation}
\Psi^+(\pm {\bf R}/r) = A \left[ \psi\left({|a \mp R| \over
2}\right) a^+ + \psi\left({|- a \mp R| \over 2}\right) b^+ \right]
\;, 
\end{equation}
$A$ being a normalization constant. The interaction operators
Eq.~(\ref{eq:int}) and (\ref{eq:H_asst.}) can now be expressed in
terms of $\Psi(\pm {\bf R}/2)$ instead of $c_{1,2}$. If the
overlap of the wave function $\psi\left({|a \mp R| \over
2}\right) \psi\left({|- a \mp R| \over 2}\right)$ is not
neglected then the assisted tunneling part of the Hamiltonian
Eq.~(\ref{eq:H_asst.}) is not invariant under the e-h
transformation but it is invariant under the left-right
exchange. Thus if the positions of the tunneling atom are not on
the lattice sites then only the commutative model shows the
combined e-h invariance. This transformation is, however,
defined differently from the one used in
Sec.~\ref{sec:newsymmetry} and the difference is in the
different momentum dependence of the transformation. We mention
at this point that the cubic tight-binding model has a very
large symmetry. Instead of the e-h transformation discussed in
this Appendix  one can also use an e-h transformation which
connects electron and hole states at the same part of the Fermi
surface, the exact analog of the symmetry transformation 
in Fig.~\ref{fig:symmetry}.\cite{Affl_Jones,Affl}

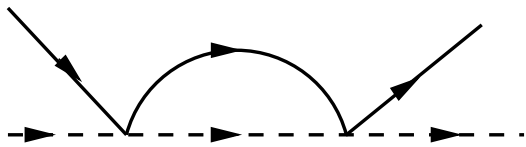
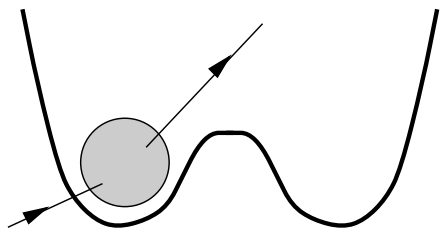
\end{document}